\documentclass[showpacs,nofootinbib,groupedaddress,superscriptaddress,preprint,showkeys]{revtex4}

\usepackage[dvips]{graphicx}
\usepackage{color} 
\usepackage{epsfig}
\usepackage{fancybox}
\usepackage{amsmath}
\usepackage{amssymb}

\allowdisplaybreaks[4]

\begin{document}

\title{Monoenergetic Neutrino Beam 
 for Long Baseline Experiments}

\author{Joe Sato}
\email{joe@phy.saitama-u.ac.jp}
\affiliation{Department of Physics, Saitama University, 
        Shimo-okubo, Sakura-ku, Saitama, 338-8570, Japan}
\affiliation{Physik-Department Technische Universit\"at M\"unchen,
       James-Franck-Strasse 85748 Garching, Germany}

\preprint{STUPP-05-180}
\preprint{TUM-HEP-581/05}

\pacs{14.60.Pq,23.40.Bw}

\keywords{long baseline experiment, neutrino oscillation,
monoenergetic neutrino, $e^-$ capture, $\beta^+$ decay
}

\begin{abstract}
In an electron capture process by a nucleus, emitted neutrinos
are monoenergetic. By making use of it,
we study how to get a completely monoenergetic neutrino beam in a
long baseline experiment.
\end{abstract}

\maketitle

\section{Introduction}
Numerous observations on neutrinos from 
the sun\cite{solar}, 
the atmosphere\cite{atm}, 
the reactor\cite{reactor},
and the accelerator\cite{K2K} suggest that
neutrinos are massive and hence there are mixings in the lepton sector.

Within the three generations, two of the mixing angles and the two mass
differences are well determined.\cite{Param} To determine these
parameters much more precisely and to observe effects from the other two
mixing parameters, $\theta_{13}$ and CP phase $\delta$, there are
several ideas proposed for next generation of neutrino oscillation
experiments.\cite{T2K,nuFact,beta-beam1,beta-beam2}

For a precision measurement, it is apparently better that we make an
experiment using neutrinos with manageable and precisely known energy.
In this respect we consider making use of a nucleus which absorbs an
electron and emits a neutrino:
\begin{eqnarray}
(Z,A)+e^-\rightarrow (Z-1,A)+\nu_e ,
\label{eq:Capture}
\end{eqnarray}
where $Z$ is the electric charge of the mother nucleus and $A$ is its
mass number.  In this case neutrinos have a line spectrum and its
energy is precisely known. Therefore by accelerating the mother nuclei
appropriately with the Lorenz boost factor $\gamma_m$, we can control
neutrino energy and make use of monoenergetic neutrinos in an
oscillation experiment.

The experimental setup is very simple. We need an accumulating ring as
usual\cite{beta-beam1} to circulate nucleus. This ring equips an
electron injection at the entrance of the decay section with its length
$X $and an apparatus for separation of nuclei and electrons at the exit
of the decay section.  The injected electrons must be tuned precisely so
that their boost factor $\gamma_e$ must be same as that of nuclei
$\gamma_m$, $\gamma_e=\gamma_m$.\footnote{When a free electron falls
into an orbit of a nucleus, a photon with energy of minding energy, $B$,
of O(1-10) keV is emitted. It gives uncertainty of momentum $B^2/M\sim
$O(1)eV for a nucleus with mass $M$ in the rest frame of the
nucleus. However it is much smaller than the nucleus momentum in the
laboratory frame so that we can safely ignore this effect. Also it takes
a time of O($10^{-9}/Z^4$) sec for an electron to fall into an orbit of
a nucleus which is much shorter than the duration within which the
nucleus stays in the decay section in the rest frame, $X/\gamma_m >
10^{-8}$ sec. Therefore we can assume that it occurs instantly.}  The
separation section at the exit must be also constructed so that it can
separate the nuclei and electron properly to circulate the nuclei till
its decay. It may be implemented by photon injection and a strong magnet.

The range of neutrino energy, $E_\nu$, in the laboratory system is given by
\begin{eqnarray}
0< E_\nu<2\gamma_m Q
\label{eq:EandQ}
\end{eqnarray}
where $Q$ is the energy difference between the mother and the daughter
nuclei and $Q\ll M$.

The appropriate energy for the experiment is derived from the baseline
length $L$ and the relevant mass square difference $\delta m^2$:
\begin{eqnarray}
 \left.\frac{\delta m^2 L}{4 E_\nu}\right|_{E_\nu=2\gamma_m Q}=P,
\label{eq:OscillationPhase}
\end{eqnarray}
where $P$ is the oscillation phase at the maximum energy of neutrino which
is determined from the physics goal. For example, if one wants to
observe the oscillation at the first maximum, then $P=\pi/2$.  From
eq.(\ref{eq:OscillationPhase}),
\begin{eqnarray}
\gamma_m= \frac{\delta m^2 L}{8P}\frac{1}{Q} .
\label{eq:OscillationPhase-gamma-and-Q}
\end{eqnarray}
Since in the rest frame of the mother nuclei, the distance between the
decay pipe and the detector is $L'\equiv L/\gamma_m$, the larger
$\gamma_m$ means the higher neutrino flux at a detector. It scales
proportionally to $\gamma_m^2$.  It means from
eq.(\ref{eq:OscillationPhase-gamma-and-Q}) that the lower $Q$ value is
better. However a lower $Q$ means, in general, a larger half-life
$\tau$. The mother nuclei should capture an electron frequently enough,
otherwise we cannot get neutrino beam of a sufficient strength. It means
\begin{eqnarray}
 \tau\gamma_m < T  \quad \Rightarrow\quad\frac{\delta m^2 L}{8PT}< Q/\tau
\label{eq:QfromT}
\end{eqnarray}
where $T$ is an appropriate time interval within which we require that
all the mother nuclei should experience the process
(\ref{eq:Capture}). Therefore, since in this kind of experiments data
are taken for several years, $T$ is of order a month or at most a
year. This requires that $\gamma_m$ should be smaller and it conflicts
with the requirement to get a higher-flux neutrino beam mentioned below
eq. (\ref{eq:OscillationPhase-gamma-and-Q}).  To satisfy both the
requirements, we have to find a nucleus which has a smaller $Q$ value and
a shorter half-life $\tau$.  In the following $\gamma_m\gg 1$ and nuclei
mass $M \gg Q$ are used to derive equations.

Here we examine the theoretical aspects of this idea in more detail.

\vspace*{0.2cm}
{\it Case (i)}
Purely monoenergetic neutrino:

As one of the first candidates we study here $^{110}_{\ 50}$Sn.
Theoretically this gives the best example for our scenario.  Its
half-life $\tau_{\rm Sn}$ is 4.11 hour. Its $J^P$ is $0^+$.  It decays
into the excited state of $^{110}_{\ 49}$In, with $1^+$ whose energy
level is 343 keV. Since the mass difference is 638 keV,
\cite{TableOfIsotopes} the energy difference between neutral
$^{110}_{\ 50}$Sn and
$^{110}_{\ 49}$In, $\Delta_{\rm Sn}$, is 295 keV, that is,
the energy of the emitted neutrino is 295 keV minus
to the binding energy.  For example, since the K shell binding energy,
$E_{\rm In}^K$ of $^{110}_{\ 49}$In is 28keV\cite{ShellTable}\footnote{
Our picture for K shell electron capture is
that a neutral mother captures its K shell electron and bears 
a neutral daughter with one K shell hall and one electron
in outer orbit. Therefore, exactly
speaking, we need to take into account the binding energy of an electron
in the outer orbit, $E_o$ which will fall into the K shell finally. This
raise the neutrino energy by amount of $E_o$, though we will omit this
here.}, the emitted neutrino energy in the rest frame of Sn, $Q_{\rm
Sn}=\Delta_{\rm Sn}-E^K_{\rm In}$ is 267 keV. \footnote{
Since an electron is captured not only from
K shell but also other orbits, there are several lines depending on from
which shell an electron is captured.
It should be
included to consider the detail.  }

Then the appropriate acceleration of $^{110}_{\ 50}$Sn is
\begin{eqnarray}
\gamma_{\rm Sn}
= 378
\left(\frac{\delta m^2}{2.5\times 10^{-3}{\rm eV}^2}\right)
\left(\frac{L}{100{\rm km}}\right)
\left(\frac{\pi/2}{P}\right).
\label{eq:GammaForSn}
\end{eqnarray}
In the rest frame of $^{110}_{\ 50}$Sn, the distance $L'_{\rm Sn}$
is given by
\begin{eqnarray}
L'_{\rm Sn}= 264 {\rm m}
\left(\frac{2.5\times 10^{-3}{\rm eV}^2}{\delta m^2}\right)
\left(\frac{P}{\pi/2}\right).
\label{eq:DistanceForSn}
\end{eqnarray}
Therefore if the "fiducial" detector radius is larger than
$264\left(\frac{P}{\pi/2}\right)$ m,  half the neutrinos goes into
the detector. Because of the reason mentioned below the theoretically most
interesting oscillation phase is $P={\pi/3}$ and hence
$264\left(\frac{P}{\pi/2}\right)=176$m. This size of a detector is not
unrealistic. Incidentally, since $\gamma_{\rm Sn}=567$, $\gamma_{\rm
Sn}\tau_{\rm Sn}=96$ days, satisfying eq. (\ref{eq:QfromT}). This
efficiency should be compared with the case of a neutrino factory or a
beta beam.  In a neutrino factory\cite{nuFact} the distance, $L'_{\mu}$
corresponding to $L'_{\rm Sn}$ is O(10) km
and hence even if the area of the detector perpendicular to the neutrino
beam is of $\left({\rm O}(100)m\right)^2$, only 0.01\% of the neutrinos are
used.  Similarly in a beta beam experiment $L'_\beta$ is O(1) km and
only 1\% of neutrinos are used.
%
%
Therefore,
%
%
even if we have by 2 orders of magnitude a smaller amount of $^{110}_{\
50}$Sn nuclei in decay than nuclei in a beta beam experiment, say
$^6_2$He, we will have same reach for the physics. That is, the
``quality factor''\cite{beta-beam1} is much better. Indeed $L'$ is
essentially the inverse of the quality factor.  Furthermore, since the
neutrino energy is much more clearly determined in this experiment, we
have better precision.

There is another interesting feature in sufficiently high $\gamma_m$
experiment.  As we have seen, almost all neutrinos go through the
detector. Therefore we have a wide range of neutrino energies and from
the detection point the neutrino energy is ``measured'' precisely.
The energy of a neutrino, which is detected at $R$ away from the center
of the beam, is easily calculated (in large $\gamma_m$ limit):
\begin{eqnarray}
 E_\nu(R)=\frac{2 \gamma_m Q}{1+R^2/L'^2}.
\label{eq:EnergyVSRadius}
\end{eqnarray}
The neutrino energy range is determined by
eq.(\ref{eq:EnergyVSRadius}) ,
\begin{eqnarray}
 \frac{2\gamma_m Q}{1+ D^2/L'^2}<E_\nu<2\gamma_m Q,
\label{eq:TrueDistance}
\end{eqnarray}
where $D$ is the ``fiducial'' detector diameter.  For example, if
$D=L'$, then the half of emitted neutrinos hits the detector and their
energy range is $\gamma_mQ\le E_\nu\le 2\gamma_mQ$.  The range of the
oscillation phase varies from $\pi/3$ to $2\pi/3$, from which we can
explore the oscillation shape around the oscillation maximum very
precisely.

For the position resolution $\delta R
(\delta R^2=2R \delta R)$, the energy
resolution is given by
\begin{eqnarray}
\left| \delta E_\nu \right|= \frac{2\gamma_m Q\delta R^2/L'^2}
{\left(1+R^2/L'^2\right)^2}\ 
\ \Rightarrow \ 
\left|\frac{\delta E_\nu}{E_\nu}\right| = \frac{\delta R^2/L'^2}
{\left(1+R^2/L'^2\right)}
.
\end{eqnarray}
In the rest frame of the mother nucleus, monoenergetic 
neutrino is emitted isotropically. In a solid angle $d\Omega$,
the number of neutrinos distribute uniformly. The solid angle
$d\Omega=2\pi\sin\theta d\theta$ corresponds to
\begin{eqnarray}
2\pi\sin\theta d\theta=\frac {4\pi}
{\left(1+R^2/L'^2\right)^2}\frac{dR^2}{L'^2}
\end{eqnarray}
and in terms of the neutrino energy
\begin{eqnarray}
d\Omega=2\pi\sin\theta d\theta= \frac{2\pi}{\gamma_m Q}\ {dE_\nu .}
\end{eqnarray}

Thus we have the neutrino beam uniformly distributed in its energy.  As a
detector can measure the energy itself, by combining energy informations
from the detection position, we can determine the neutrino
energy very precisely.  This specific feature in a beta-capture beam arises
from the fact that neutrinos are monoenergetic in the rest
frame of the mother nucleus.

In table. \ref{tb:ScenarioI}, we list nucleus candidates for this case (i).
\begin{table*}
\begin{tabular}[t]{|c|c|c|c|c|c|c|}
\hline
 \makebox[30mm] {Mother, $E^K$\cite{ShellTable}} &
\makebox[30mm] {Daughter, $E^K$\cite{ShellTable}}
 & \makebox[20mm]{ $\Delta$
\cite{TableOfIsotopes}}
&$\tau$\cite{TableOfIsotopes}
&$\gamma_m$&$\tau\gamma_m$&Detector Size\\
\hline\hline
$^{110}_{\ 50}$Sn ,29 & $^{110}_{\ 49}$In$^*$ [343], 28 
& 295&\makebox[15mm] {4.11 h}&\makebox[15mm] {567}
&\makebox[15mm]{97 d}&176 m\\
\hline
$^{111}_{\ 49}$In, 28 &
$^{111}_{\ 48}$Cd$^*$[417], 27 &
 449&2.80 d&359&1005 d&278m\\
\hline
\end{tabular}
\caption{Nucleus candidates for case (i). $\gamma_m$ is determined by
$P=\pi/3$ for a detector at $L=100$km and $\delta
m^2=2.5\times10^{-3}$eV$^2$ using
eq.(\ref{eq:OscillationPhase-gamma-and-Q}). The energy unit is
keV. N$^*$[E] means the excited state of the nucleus N with energy
E[keV].  ``Detector Size'' indicates the radius within which a half of
the emitted neutrinos are included at the detector distance, see
eq.(\ref{eq:TrueDistance}).}  \label{tb:ScenarioI}
\end{table*}

\vspace*{0.2cm}
{\it case (ii)}
Monoenergetic neutrino and Continuous energy neutrino:

Next we consider the nuclei $^{48}_{24}$Cr. 
It decays into an excited state of $^{48}_{23}$V
whose energy level is 420keV. The mass difference is 1659keV
and $\Delta_{\rm Cr}$ is 1239 MeV.
The half-life is 21.56 hours.\cite{TableOfIsotopes}
K shell binding energy, $E_{\rm V}^K$, of the daughter nucleus $^{48}_{23}$V
is 5.465 keV\cite{ShellTable}.
Since $\ Q_{\rm Cr}=\Delta_{\rm Cr}-E_{\rm V}^K
$ is larger than 2$m_e$, twice of the electron mass, 
it not only captures an electron but also emits a positron:
\begin{eqnarray}
 ^{48}_{24}{\rm Cr}+e^-\rightarrow ^{48}_{23}{\rm V}+\nu_e\ \& 
\ ^{48}_{24}{\rm Cr}\rightarrow ^{48}_{23}{\rm V}+e^++\nu_e .
\end{eqnarray}
Assuming that there are 2 K shell electrons in the mother
nucleus $^{48}_{24}$Cr,
the rate for the capture process, $\Gamma_c$, is proportional to
\cite{FY}
\begin{eqnarray}
 \Gamma_c\propto 2 \pi
\left\{(Q_{\rm Cr})/m_e\right\}^2 (\alpha Z)^3
=0.196 .
\label{eq:GammaC}
\end{eqnarray}
Here $m_e$ is the electron mass.  The rate for positron emission ,
$\Gamma_{e^+}$, is proportional to \cite{FY}
\begin{eqnarray}
 \Gamma_{e^+}\propto
\int^{w_0}_1 x \sqrt{x^2-1} (w_0-x)^2F(x,Z) dx
=0.004,
\label{eq:GammaE}\\
F(x,Z)=
2(1+\gamma)\left\{2pr\right\}^{2\gamma-2}\hspace{-0.3cm}
\exp (-\pi \nu)\frac{\left|\Gamma(\gamma-i\nu)\right|^2}
{\left[\Gamma(2\gamma+1)\right]^2}.
\end{eqnarray}
Here $F(x,Z)$ is the Fermi function ($\gamma\equiv(1-\alpha Z)^{1/2},
\nu\equiv\alpha Z x/p, p=\sqrt{x^2-1}$, $\alpha$ the fine structure
constant =1/137 and $r$ the radius for a nucleus in units of $m_e^{-1}$)
\footnote{ For numerical calculation we take $r=10^{-3}$. However the
numerical results here does not depend on $r$ within a few \% accuracy.}
and $w_0=(\Delta_{\rm Cr}-m_e)/m_e$ is the maximum positron energy
scaled by an electron mass.  Thus the electron capture process is
dominant (98.0\%) and hence a neutrino beam with well-controlled energy
is available.

\begin{table*}
\begin{tabular}[t]{|c|c|c|c|c|c|ccc|}
\hline
 \makebox[30mm] {Mother, $E^K$\cite{ShellTable}} &
\makebox[30mm] {Daughter, $E^K$\cite{ShellTable}
} & \makebox[20mm]{ $\Delta$
\cite{TableOfIsotopes}}
&$\tau$\cite{TableOfIsotopes}
&$\gamma_m$&$\tau\gamma_m$&EC&:&$e^+$emission\\
\hline\hline
$^{18}_{\ 9}$F, 0.7 & $^{18}_{\ 8}$O ,0.5
& 1656&\makebox[15mm] {110 m}&\makebox[15mm] {123}
&\makebox[15mm]{4.65 d}&3.4 &:& 96.6\\
\hline
$^{48}_{24}$Cr, 6 &
$^{48}_{23}$V$^*$[420], 5 &
 1239&21.56 h&82&74 d&98.0 &:& 2.0\\
\hline
$^{111}_{\ 50}$Sn ,29& $^{111}_{\ 49}$In, 28&
 2445&35.3 m&42&24.7 h&40.5 &:& 59.5\\ 
\hline
$^{113}_{\ 50}$Sn$^*$[77], 29& $^{113}_{\ 49}$In, 28&
 1113&21.4 m&93&33.2 h&100&:&0\\ 
\hline
\end{tabular}
\caption{Candidate
Nuclei for case (ii). $\gamma_m$ is determined by $P=\pi/2$
instead $\pi/3$. 
The last column,
the ratio of the electron capture and the positron emission, is
calculated by using eq.(\ref{eq:GammaC}) and eq.(\ref{eq:GammaE}).
}
\label{tb:ScenarioII}
\end{table*}

In table \ref{tb:ScenarioII} we list other examples of nuclei
which have still lower $Q$ and shorter $\tau$\cite{TableOfIsotopes}.
$^{18}_{\ 9}$F dominantly decays by positron emission while
$^{48}_{24}$Cr and $^{113}_{\ 50}$Sn$^*$ 
almost capture an electron to bear their daughter nucleus.

Since $Q_{\rm Cr}$ is higher than the previous case,
the appropriate $\gamma_{\rm Cr}$
is lower and hence the quality factor is worse than the previous case.
Namely, we need to prepare much more $^{48}_{24}$Cr nuclei than
$^{110}_{\ 50}$Sn:
\begin{eqnarray}
\gamma_{\rm Cr}
= 82
\left(\frac{\delta m^2}{2.50\times 10^{-3}{\rm eV}^2}\right)
\left(\frac{L}{100{\rm km}}\right)
\left(\frac{\pi/2}{P}\right).
\label{eq:GammaForCr}
\end{eqnarray}
which means that the neutrino at the
detector is completely monoenergetic. There is essentially no position
dependence of neutrino energy at the detector.

We also cannot explore the energy dependence of the oscillation
simultaneously as previously discussed. However, this problem may be
solved by the use of continuous neutrino associated with positron
emission. We can control the boost factor $\gamma_m$ very well and hence
the highest neutrino energy at a detector is completely determined from
it. This offers very accurate calibration for neutrino energy. 
Furthermore, the energy of the line spectrum and that of
the continuous one are clearly separated and simultaneous observation
of two distinct energy region gives a useful information on Unitarity
triangle\cite{JS}. Thus
having a line and a continuous spectrum simultaneously, we may get
better oscillation parameter reach.

\vspace{0.5cm}

We study how to control neutrino energy in oscillation experiments
better than currently discussed ideas. By electron capture, a nucleus
emits a monoenergetic neutrino. Therefore by accelerating the mother
nuclei, we can get a well-controlled neutrino beam. To achieve 100 \%
electron capture rate, we need to use a nucleus with a low $Q$ value,
lower than 2$m_e$.  In general, such a nucleus has a long half-life.
Furthermore, since we accelerate it with significantly large boost
factor $\gamma_m$, it becomes almost stable.  Though this easily
conflicts with the fact that the nucleus must decay within sufficiently
short interval (see eq.(\ref{eq:QfromT})), there are
several candidates listed in table \ref{tb:ScenarioII}. With these
nuclei, we can control neutrino energy.  Since $\gamma_m$ is very large,
a neutrino beam is so well concentrated in the forward direction that
almost all neutrinos can be used for oscillation experiments. It reduces
significantly the necessary number of the mother nuclei. As a result of
such a high $\gamma_m$, in principle, we don't need to measure the
neutrino energy at a detector since by measuring the detected position
we can calculate its energy and hence simultaneously we can observe the
energy dependence of the oscillation.

Theoretically there are only advantages but these nuclei are so
heavy that it is very energy consuming to accelerate them to an ideal
$\gamma_m$. Also it may be hard to get enough nuclei even if the
required number of nuclei is significantly small.  As a compromise, we
also study nuclei with a higher $Q$ value.  Those nuclei not only capture
an electron but also emit a positron.  From the latter process neutrinos with
continuous spectrum are emitted.  Furthermore as $Q$ is higher,
$\gamma_m$ must be smaller.  These facts spoil some of the good features
mentioned above.  However since we have neutrinos with a line spectrum and
a continuous spectrum simultaneously, we may get another good feature for
this kind of beams. 

In this kind of a beta-capture beam, we can produce only $\nu_e$ beam.
To study CP violation we need $\bar\nu_e$\cite{AS} or
$\nu_\mu$\cite{AKS} beam.  On the contrary to $e^-$ capture case, since
$e^+$ cannot be bound by a nuclei, it is almost impossible to have a
sufficiently strong $\bar\nu_e$ beam. Instead we can make use of $\mu$
capture to get monoenergetic $\nu_\mu$ beam, though since the mass of
$\mu$ is very high, emitted neutrinos have a high energy.  We must find
a nucleus whose daughter has a mass higher than that of the mother by
O($\mu$) mass so that the energy of $\nu_\mu$ in the rest frame of the
mother nucleus is sufficiently low.

Apart from the idea to make use of $e^-$ capture,
a nucleus $^{18}_{\ 9}$F should be considered as
the $\beta$ beam source more seriously.
Note that in an ideal circumstance, with static strong magnetic
large circulating ring, etc.,
we do not need any power supply to
maintain current by nuclei.
Therefore the mother nuclei do not have to decay ``immediately''
Since $\Delta_{\rm F}$ is 1655.5 keV while $\Delta_{\rm Ne}$
is 4446 keV, we have much better``quality factor''
than $^{18}_{10}$Ne.
$^{18}_{\ 9}$F is used
for medical check, Positron Emission Tomography (PET).
They are made within one hour about O$(10^{10})$ Bq, about
$10^{14}$ nuclei per hour $\simeq 10^{18}$ nuclei per year
even in a medical check. We can use a much larger amount
of such a nuclei much more easily than $^{18}_{10}$Ne. 
Similarly we need to reconsider a candidate nuclei for $\bar\nu_e$
source with lower $Q$ than $^{6}_{2}$He, e.g. $^{31}_{14}$Si.

\noindent {\large \it Acknowledgments}\\ The author thanks T. Ota for
useful comment.  The author also thanks T. Higuchi and M. Ishiduki about
information on PET.  He is supported by the Grant-in-Aid for Scientific
Research on Priority Area No.16038202 and 14740168.


\newcommand{\Journal}[4]{{ #1} {\bf #2} {(#3)} {#4}}
\newcommand{\APJ}{Ap. J.}
\newcommand{\CJP}{Can. J. Phys.}
\newcommand{\EPJ}{Eur. Phys. J.}
\newcommand{\MPL}{Mod. Phys. Lett.}
\newcommand{\NC}{Nuovo Cimento}
\newcommand{\NP}{Nucl. Phys.}
\newcommand{\PL}{Phys. Lett.}
\newcommand{\PR}{Phys. Rev.}
\newcommand{\PRep}{Phys. Rep.}
\newcommand{\PRL}{Phys. Rev. Lett.}
\newcommand{\PTP}{Prog. Theor. Phys.}
\newcommand{\SJNP}{Sov. J. Nucl. Phys.}
\newcommand{\ZP}{Z. Phys.}
\newcommand{\EUR}{Eur. Phys. J.}


\end{document}